\documentclass[preprint,showpacs,preprintnumbers,amsmath,amssymb]{revtex4}

\usepackage{amsmath}
\usepackage{amssymb}
\usepackage{graphicx}
\usepackage{epsfig,subfigure}

\newcommand {\nn}    {\nonumber}

\begin{document}

\title{Zero Modes of Matter Fields on Scalar Flat Thick Branes
}

\author{Li-Jie Zhang \footnote{E-mail: lijzhang@shu.edu.cn},
        Guo-Hong Yang \footnote{E-mail: ghyang@shu.edu.cn}}

\affiliation{
   Department of Physics, Shanghai University,
       Shanghai 200444, China \\
   }

\begin{abstract}
Zero modes of various matters with spin 0, 1 and 1/2 on a class of
scalar flat thick branes are discussed in this paper. We show that
scalar field with spin 0 is localized on all thick branes without
additional condition, while spin 1 vector field is not localized.
In addition, for spin 1/2 fermionic field, the zero mode is
localized on the branes under certain conditions.
\end{abstract}

 \pacs{11.10.Kk, 04.50.-h
 }





\maketitle

\section{Introduction}

In the 1920's Kaluza and Klein came up with the Compactification
theory. In their original work it was shown that if we start with
a theory of general relativity in 5-spacetime dimensions and then
curl up one of the dimensions into a circle we end up with a
4-dimensional theory of general relativity plus electromagnetism.
More recently, at least since about 1980, the Kaluza-Klein theory
has been revived in attempts to derive supergravity theories.
Nowadays it has been acknowledged that our four dimensional world
is a three-brane embedded in a higher dimensional space-time with
non-factorizable warped
geometry.\cite{braneworld1,braneworld2,braneworld3,braneworld4} It
means that we are free from the moduli stabilization problem in
the sense that the internal manifold is noncompact and does not
need to be compactified to the Planck scale any more.

The most popular model in this field is the so-called RS
model\cite{RS1,RS2} which consists of a flat brane embedded in an
AdS5 bulk. The particular idea of the model was the inclusion of
tension of the brane, describing the energy density per volume in
three-dimensional space. The researchers focused on models with
only one extra dimension, but the basic concept can be generalized
to an arbitrary higher-dimensional spacetime. Positive result
suggests Randall-Sundrum extra dimensions are correct, and we
could extend RS brane worlds in principle. According to the model,
spin 0 field is localized on a brane with positive tension, spin
1/2 and 3/2 fields are localized not on a brane with positive
tension but on a brane with negative tension.\cite{localization1}

The key ingredient for realizing the brane world idea is the
localization of various bulk fields on a brane by a natural
mechanism. Gravity is known to be the unique interaction having
universal coupling with all matter fields. Considering RS model in
the thin brane limit, thick brane scenarios based on gravity
coupled to scalars have been constructed recently\cite{f,g1,g2}.
With these models, one can obtain branes naturally without
introducing them by hand in the action of the theory \cite{f}. And
some work has been made to generalize the models which based on
gravity coupled to scalars
\cite{Arias,Bazeia2004,Barbosa1,Barbosa2,BazeiaJHEP2006,Liu0708}.
Then it is of great interest to extend the investigation to the
localization of various matters on the thick branes.

The organization of the paper is as follows: The first part of
this paper will review the thick brane worlds arising from scalar
fields, then we will discuss in detail the localization of various
matters with spin 0, 1 and 1/2.  One may ponder in this section
whether all the fields can be localized on various thick branes.
Finally, in Section 4, we summarize what have been achieved in the
paper, and give some further discussions.

\section{Review of thick brane worlds arising from scalar fields}
\label{SecModel}

The model that we investigate is described by a theory of
five-dimensional gravity coupled to scalar fields governed by the
following action
\begin{equation}\label{model}
S=\int\,d^4xdy\;{\sqrt{|g|}\;\left(\frac14\,R+{\cal
L}(\phi,\partial_i\phi)\right)},
\end{equation}
where $\phi$ stands for a real scalar field and we are using
${4\pi G}=1.$ These theories with one or many scalar fields can
simulate true five-dimensional supergravity theories under certain
consistent truncations. The line element $ds^2_5$ of the
five-dimensional space-time can be written as
\begin{equation}\label{metric}
 ds^2_5=g_{MN}dx^M dx^N
 =e^{2 A}ds^2_4+dy^2,
\end{equation}
where $M,N=0,1,...,4$. Also, $ds^2_4$ represents the line element
of the four-dimensional space-time, which can have the form
\begin{eqnarray}
 ds^2_4&=&-dt^2+e^{2\sqrt{\Lambda}\;t}(dx_1^2+dx_2^2+dx_3^2),\\
 ds^2_4&=&e^{-2\sqrt{\Lambda}\;x_3}(-dt^2+dx_1^2+dx_2^2)+dx_3^2,
\end{eqnarray}
for dS and AdS geometry, respectively. Here $\mathrm{e}^{2A}$ is
the warp factor and $\Lambda$ represents the cosmological constant
of the four-dimensional space-time; the limit $\Lambda\to0$ leads
to the line element
\begin{equation}
ds^2_5=e^{2A}\eta_{\mu\nu}dx^\mu dx^\nu+dy^2,
\end{equation}
where $\eta_{\mu\nu}$ describes Minkowski geometry. The scalar
field dynamics is governed by the Lagrangian density
\begin{equation}
{\cal L}=-\frac12\,g^{MN}\,\partial_M\phi\,\partial_N\phi-V,
\end{equation}
where $V=V(\phi)$ represents the potential, which specifies the
model to be considered.

The standard scenario is to suppose that both $A$ and $\phi$ are
static, depending only on the extra dimension, that is we can set
$A=A(y)$ and $\phi=\phi(y)$. The example for flat
branes\cite{ablPRD2006} is given by
\begin{eqnarray}
V=\frac{1}{6} \phi ^{2-\frac{2}{p}} \left[3 \left(\phi
^{\frac{2}{p}}-1\right)^2-8 p^2 \phi ^2 \left(\frac{1}{2
p-1}-\frac{\phi ^{\frac{2}{p}}}{2 p+1}\right)^2\right],\nn
\end{eqnarray}
where $p$ is odd integer. The case $p=1$ is special, and
reproduces the $\phi^4$ model in flat spacetime. This model was
first introduced in \cite{bmmPRL2003}, and it was recently
considered within the braneworld context in \cite{Bazeia2004}. The
scalar field and the warp factor are given by
\begin{eqnarray}
 \phi&=&\tanh^p(\frac{y}{p}),\label{phi1}\\
 A&=&-\frac{1}{3}\frac{p}{2p+1}\tanh^{2p}(\frac{y}{p})\nn \\
     &&+\frac{4}{3}\frac{p^2}{1-4p^2}
     \left[\ln\cosh(\frac{y}{p})
     -\sum_{n=1}^{p-1}\frac{1}{2n}\tanh^{2n}(\frac{y}{p})\right].~~~
 \label{A1}
\end{eqnarray}
The energy density of the scalar matter is
\begin{equation}
 T_{00}(y) \propto -3 \mathrm{e}^{2A} \left( 2A'^2+A'' \right).
 \label{density1}
\end{equation}
For $p=1$, this function has two negative minima and a positive
maximum at $y=0$, and finally it vanishes asymptotically.
The fact can be compared with Randall-Sundrum thin brane case,
where one of the branes has a positive brane tension meanwhile the
second brane has a negative one \cite{RS1,RS2,rs}. For the case of
$p>1$, however, this function has two negative minima and two
positive maxima at $y\neq0$, and finally it vanishes also
asymptotically. Note that, for this case, the energy density at
$y=0$ is zero.



\section{Localization of various matter fields}
\label{SecLocalize}

In this section, we focus attention on the problems of
localization of various bulk fields with spin 0, 1 and 1/2. It is
necessary to point out that the solutions given in previous
section remain valid even in the presence of bulk fields since
various bulk fields considered below make little contribution to
the bulk energy. Now we discuss the situation of various matter
fields in detail.

\subsection{Spin 0 scalar field}

For a real scalar field on the thick branes, we first investigate
how to localize it in the backgrounds (\ref{phi1}, \ref{A1}). Let
us consider the action of a massless real scalar coupled to
gravity
\begin{eqnarray}
S_0 = - \frac{1}{2} \int d^5 x  \sqrt{-g}\; g^{M N}
\partial_M \Phi \partial_N \Phi,
\label{scalarAction}
\end{eqnarray}
from which the equation of motion can be derived
\begin{eqnarray}
\frac{1}{\sqrt{-g}} \partial_M (\sqrt{-g} g^{M N} \partial_N \Phi)
= 0. \label{scalarEOM}
\end{eqnarray}
For simplicity, we define $P(y)=e^{2A(y)}$. Then the background
metric is determined by
\begin{eqnarray}
ds_5^2=P(y)\eta_{\mu\nu}dx^\mu dx^\nu+dy^2, \label{MetricUsePQ}
\end{eqnarray}
and the equation of motion (\ref{scalarEOM}) becomes
\begin{eqnarray}
P\eta^{\mu\nu} \partial_\mu \partial_\nu \Phi +
\partial_y (P^2 \partial_y \Phi) = 0. \label{36}
\end{eqnarray}
Making the following decomposition
\begin{eqnarray}
\Phi(x,y) = \phi(x) \chi(y) = \phi(x) \sum_{m} \chi_m(y),
\label{37}
\end{eqnarray}
and demanding $\phi(x)$ satisfies the massless 4-dimensional
Klein-Gordon equation $\eta^{\mu\nu} \partial_\mu \partial_\nu
\phi(x) = 0 $, we obtain the zero-mass constant solution of Eq.
(\ref{36}), i.e. $\chi(y) = \chi_0 = constant$.

Now we show that this constant mode is localized on the thick
branes arising from scalar kink (\ref{phi1}). The condition for
having localized 4-dimensional scalar field is that $\chi(y) =
\chi_0$ is normalizable, which is equivalent to the condition that
the ``coupling" constant is nonvanishing. Substituting the zero
mode into the starting action (\ref{scalarAction}), we get
\begin{eqnarray}
S_0^{(0)}
 = - \frac{1}{2} \chi_0^2 \int_{-\infty}^{\infty} dy P \int d^4 x
\eta^{\mu\nu}
\partial_\mu \phi
\partial_\nu \phi. \label{40}
\end{eqnarray}
Hence, in order to localize $4$-dimensional scalar field on the
thick branes one requires that the integral $I_0$, which is
defined by
\begin{eqnarray}
I_0 &=& \int_{-\infty}^{\infty} dy P  , \label{I0}
\end{eqnarray}
should be finite. Noting that $\tanh^{2n}(y/p)\leq 1$ for
arbitrary finite $y$ and positive $n$, we have
\begin{eqnarray}
 I_0 &=& \int_{-\infty}^{\infty} dy P
      < \int_{-\infty}^{\infty} dy P_{\infty}
       \label{I0a}
\end{eqnarray}
is finite for arbitrary positive odd $p$, where
\begin{eqnarray}
P_{\infty}\equiv
 \left(\cosh\frac{y}{p}\right)^{-\frac{8p^2}{3(4p^2-1)}}
 e^{\frac{2}{3}\frac{p}{2p+1}
        \left( \frac{2p}{2p-1} \sum_{n=1}^{p-1}\frac{1}{n} - 1
        \right)}.\nn
\end{eqnarray}
For examples, we have
\begin{eqnarray}
 I_0 &<& \frac{e^{-2/9}\sqrt{\pi}~\Gamma(4/9)}{\Gamma(17/18)}
 =1.5 \sqrt{\pi}
 ~~~~~~~  \text{for}  ~ p=1, \nn\\
 I_0 &<&
 \frac{3e^{8/35}\sqrt{\pi}~\Gamma(12/35)}{\Gamma(59/70)}=8.8
 \sqrt{\pi}
 ~~~  \text{for}  ~ p=3. \nn
\end{eqnarray}

From above analysis, we conclude that the spin 0 scalar field can
be localized on thick branes without additional conditions.

\subsection{Spin 1 vector field}

Next we turn our attention to spin 1 vector field. Now the action of
$U(1)$ vector field is considered as the following:
\begin{eqnarray}
S_1 = - \frac{1}{4} \int d^5 x \sqrt{-g} g^{M N} g^{R S} F_{MR}
F_{NS}, \label{44}
\end{eqnarray}
where $F_{MN} = \partial_M A_N - \partial_N A_M$. And the equation
motion can be obtained easily by using this action, i.e.
\begin{eqnarray}
\frac{1}{\sqrt{-g}} \partial_M (\sqrt{-g} g^{M N} g^{R S} F_{NS})
= 0.
\end{eqnarray}
From the background geometry (\ref{MetricUsePQ}), this equation is
reduced to
\begin{eqnarray}
 \eta^{\mu\nu} \partial_\mu F_{\nu y} &=& 0, \\
 \partial^\mu F_{\mu \nu}
      +  \partial_y \left(P F_{y \nu}\right)&=& 0.
\label{46}
\end{eqnarray}
By decomposing the vector field as
\begin{eqnarray}
A_{\mu}(x,y) = a_\mu(x) \sum_{m} \rho_m(y), ~~
A_y(x,y) = a_y(x) \sum_{m} \rho_m(y), \nn
\end{eqnarray}
it is easy to see that there is the constant solution $\rho_m(y) =
\rho_0 =$ constant and $a_y =$ constant if we use $\partial^\mu
f_{\mu\nu} = 0$ with the definition of $f_{\mu\nu} =
\partial_\mu a_\nu - \partial_\nu a_\mu$.

Now let us substitute this constant solution into the action
(\ref{44}) in order to see whether the solution is a normalizable
one or not. It turns out that the action is reduced to
\begin{eqnarray}
 S_1^{(0)} 
     = - \frac{1}{4} \rho_0^2  \int_{-\infty}^{\infty} dy
           \int d^4 x \eta^{\mu\nu} \eta^{\alpha\beta}
                      f_{\mu\alpha} f_{\nu\beta}. \label{48}
\end{eqnarray}
Here we define $I_1$ by
\begin{eqnarray}
I_1 &=& \int_{-\infty}^{\infty} dy ,
\end{eqnarray}
then the condition of having localized 4-dimensional vector field
on the branes requires $I_1$ to be finite. Obviously, the integral
is divergent, which shows that the vector field can not be
localized on the thick brane. This conclusion can be compared with
the work of others. It was shown in the Randall-Sundrum model in
$AdS_5$ space that spin 1 vector field is not localized neither on
a brane with positive tension nor on a brane with negative tension
so the Dvali-Shifman mechanism \cite{G. Dvali} must be considered
for the vector field localization \cite{B. Bajc}.

\subsection{Spin 1/2 fermionic field}

For spin 1/2 fermionic field, it is well known that fermions can
not be localized in Randall-Sundrum branes\cite{RS2}. Melfo {\em
et al } studied the localization of fermions on various different
scalar thick branes recently\cite{S. Randjbar-Daemi,S.L.
Dubovsky,A. Melfo}, and had shown that only one massless chiral
mode is localized in double walls and branes interpolating between
different $AdS_5$ spacetimes whenever the wall thickness is keep
finite, while Chiral fermionic modes can not be localized in
$dS_4$ walls embedded in a $M_5$ spacetime. It is therefore of
interest to investigate the possibility that fermion confinement,
being directly dependent on the scalar field solution and not only
on the spacetime metric, can be affected by the internal structure
on the thick brane. We can shown that, for the case of no Yukawa
coupling, there is no bound states for both left and right chiral
fermions. Hence, for the massless chiral fermion localization,
there must be some kind of Yukawa coupling. In this subsection
localization of a spin 1/2 fermionic field on the thick branes
will be investigated.

The discussions begin with the Dirac action of a massless spin 1/2
fermion coupled to gravity and scalar
\begin{eqnarray}
S_{1/2} = \int d^5 x \sqrt{-g} (\bar{\Psi}  \Gamma^M D_M
\Psi+\lambda\bar\Psi\phi\Psi), \label{DiracAction}
\end{eqnarray}
from which the equation of motion is given by
\begin{eqnarray}
\big(\Gamma^M  ( \partial_M + \omega_M )+\lambda \phi \big)
\Psi&=& 0, \label{DiracEq1}
\end{eqnarray}
where $\omega_M= \frac{1}{4} \omega_M^{\bar{M} \bar{N}}
\gamma_{\bar{M}} \gamma_{\bar{N}}$ is the spin connection with
$\bar{M}, \bar{N}, \cdots$ denoting the local Lorentz indices,
$\Gamma^M$ and $\gamma^{\bar{M}}$ are the curved gamma matrices
and the flat gamma ones respectively, and have the relations
$\Gamma^M = e^M _{\bar{M}} \gamma^{\bar{M}}$ with $e_M ^{\bar{M}}$
being the vielbein.
The  non-vanishing components of $\omega_M$ are
\begin{eqnarray}
  \omega_\mu = \frac{P'}{4\sqrt{P}} \gamma_\mu \gamma_y. \label{eq4}
\end{eqnarray}
And the Dirac equation (\ref{DiracEq1}) then becomes
\begin{equation}
 \left\{ P^{-\frac{1}{2}} \gamma^{\mu} \partial_{\mu}
        +\gamma^y \partial_y
        +P'P^{-1} \gamma^y+\lambda\phi
 \right \} \Psi=0, \label{DiracEq2}
\end{equation}
where $\gamma^{\mu} \partial_{\mu}$ is the Dirac operator on the
brane. We are now ready to study the above Dirac equation for
5-dimensional fluctuations, and write it in terms of 4-dimensional
effective fields. From the equation of motion (\ref{DiracEq2}), we
will search for the solutions of the chiral decomposition
\begin{equation}
 \Psi_{\pm} = \psi_{\pm}(x) \alpha_{\pm}(y), \label{Psixy}
\end{equation}
where $\psi_{\pm}(x)$ satisfies the massless 4-dimensional Dirac
equation. The equation for the chiral modes becomes
\begin{eqnarray}
  \left\{ \gamma^{\mu} \partial_{\mu}
        +P^{\frac{1}{2}} \left(\gamma^y \partial_y
        +P'P^{-1} \gamma^y+\lambda\phi \right)
 \right \} \Psi_{\pm}=0.  \label{Psixy2}
\end{eqnarray}

Setting $\gamma^y \Psi_{\pm} = \pm\Psi_{\pm}$, namely assume
$\Psi_{+}$ is left chiral fermion $\gamma^y \Psi_{+} = \Psi_{+}$,
$\Psi_{-}$ is right chiral fermion $\gamma^y \Psi_{-} =-\Psi_{-}$,
the new form of above equation can be obtained. For zero modes,
corresponding to 4D massless fermions, we have $\gamma^{\mu}
\partial_{\mu}\psi_{\pm}(x)=0$, and
\begin{eqnarray}
  \bigg(\pm\partial_y \pm \frac{P'}{P} +\lambda\phi \bigg)\alpha_{\pm}(y)
  =0.
\end{eqnarray}
So the solutions of zero modes are
\begin{eqnarray}
\Psi_{\pm} =c_0\exp \bigg\{- \int^y
\bigg(\frac{P'(\bar{y})}{P(\bar{y})} \pm \lambda\phi(\bar{y})
\bigg) d\bar{y} \bigg\} ~\psi_{\pm}(x). \label{ZeroMode}
\end{eqnarray}

Let us analyze whether this zero mode can be localized on the
branes under certain conditions. Substituting (\ref{ZeroMode})
into the Dirac action (\ref{DiracAction}), we obtain
\begin{eqnarray}
 S^{(0)}_{1/2}=I_{1/2}
   \int d^4 x ~ {\overline{\psi}_{\pm}}
         \gamma^{\mu} \partial_{\mu} \psi_{\pm},
  \label{effDiracAction}
\end{eqnarray}
where $I_{1/2}$ is defined as
\begin{eqnarray}
I_{1/2}=c_0^2 \int_{-\infty}^{\infty} dy P^{-\frac{1}{2}} e^{ -2
\int^y \pm \lambda \phi(\bar{y}) d\bar{y} }.  \label{I12}
\end{eqnarray}
In order to localize spin 1/2 fermion in this framework, the
integral (\ref{I12}) should be finite. Using the configuration
(\ref{phi1}, \ref{A1}) and $P(y)=e^{2A(y)}$, the requirement of
convergence for integral is easily discussed. For example, for
$p=1$, substituting $P = {\cosh}^{-\frac{8}{9}}y
\;\exp({-\frac{2}{9} \tanh^2 y})$ and $\phi ={\tanh}\;y $ into
(\ref{I12}), we get
\begin{eqnarray}
I_{1/2}
= c_0^2 \int_{-\infty}^{\infty} dy~
   \cosh^{\frac{4}{9} - 2(\pm\lambda)} y~
   \exp\left({\frac{1}{9} \tanh^2 y}\right).\label{I12+}
\end{eqnarray}
Noting that for $\lambda>\frac{2}{9}$, we have
\begin{eqnarray}
 \cosh^{(\frac{4}{9}- 2\lambda)} y|_{y\rightarrow\pm\infty}=0,~~
 \exp(\frac{1}{9}\tanh^2 y)|_{y\rightarrow\pm\infty}=
   e^{\frac{1}{9}}.\nn
\end{eqnarray}
Now one can check that the integral (\ref{I12+}) is finite when
$\lambda>\frac{2}{9}$ for left chiral fermion. For right chiral
fermion, the localization condition is $\lambda<-\frac{2}{9}$. For
$p=3$, using the similar method, we can also get the localization
condition for left chiral fermion is $\lambda>\frac{2}{35}$.

Based on above analysis, the requirement of convergence for
integral is studied, and the discussions end with the conclusion
that the spin 1/2 field can be localized with some conditions
satisfied.

\section{Summary}

From the viewpoint of field theory, the possibility of localizing
various matter fields on thick branes has been investigated in
this paper. The analytical study shows that spin 0 field is
localized on thick branes without additional condition. But the
result of spin 1 field is the same as the case of RS model in
$AdS_5$ space, in which vector field is not localized neither on a
brane with positive tension nor on a brane with negative tension
so the Dvali-Shifman mechanism should be considered for the vector
field localization. Since localization of fermions on branes
requires other interactions but gravity, we show that massless
chiral fermionic modes can be localized on the thick branes,
coupled through a Yukawa term. There are still some other
backgrounds, for example, vortex
background\cite{WangMPLA2005,LiuNPB2007,LiuVortex}, could be
considered besides gauge field and
supergravity\cite{LiuJHEP2007,DuanGauge,Mario}. The topological
vortex coupled to fermions may lead to chiral fermion zero
modes\cite{JackiwRossiNPB1981}. Usually the number of the zero
modes coincides with the topological number, that is, with the
magnetic flux of the vortex. Under these backgrounds, more
extensive work can be carried out in the future.

\section*{Acknowledgement}
We are really grateful to Yu-Xiao Liu for his valuable
recommendations and discussion. This work was supported by
Shanghai Education Commission, the Innovation Foundation of
Shanghai University and Shanghai Leading Academic Discipline
Project (S30105).


\begin{thebibliography}{99}

\bibitem{braneworld1}
 R. Gregory,
    Phys. Rev. Lett. \textbf{84}, 2564 (2000).

\bibitem{braneworld2}
 I. Oda,
    Phys. Lett. \textbf{B 496}, 113 (2000).

\bibitem{braneworld3}
 R . Koley and S. Kar,
     Class. Quant. Grav. \textbf{24}, 79 (2007).

\bibitem{braneworld4}
 M. Gogberashvili,
    Int. J. Mod. Phys. \textbf{11}, 1635 (2002).

\bibitem{RS1}
 L. Randall and R. Sundrum,
    Phys. Rev. Lett. \textbf{83}, 3370 (1999).

\bibitem{RS2}
 L. Randall and R. Sundrum,
    Phys. Rev. Lett. \textbf{83}, 4690 (1999).

\bibitem{localization1}
 Y. Grossman and M. Neubert,
    Phys. Lett. \textbf{B 474}, 361 (2000).

\bibitem{f}
 O. DeWolfe, D.Z. Freedman, S.S. Gubser and A. Karch,
    Phys. Rev. \textbf{D 62}, 046008 (2000).

\bibitem{g1}
 M. Gremm,
    Phys. Lett. \textbf{B 478}, 434 (2000).

\bibitem{g2}
 M. Gremm,
    Phys. Rev. \textbf{D 62}, 044017 (2000).

\bibitem{Arias}
O. Arias, R. Cardenas and I. Quiros,
    Nucl. Phys. \textbf{B 643}, 187 (2002).

\bibitem{Bazeia2004}
 D. Bazeia, C. Furtado and A.R. Gomes, JCAP \textbf{0402}, 002 (2004).

\bibitem{Barbosa1}
 N. Barbosa-Cendejas and A. Herrera-Aguilar, JHEP \textbf{0510}, 101 (2005).

\bibitem{Barbosa2}
 N. Barbosa-Cendejas and A. Herrera-Aguilar, Phys. Rev. \textbf{D 73}, 084022 (2006).

\bibitem{BazeiaJHEP2006}
 D. Bazeia, F.A. Brito and L. Losano, JHEP \textbf{0611}, 064 (2006).

\bibitem{Liu0708}
 Y.X. Liu, X.H. Zhang, L.D. Zhang and Y.S. Duan, JHEP \textbf{0802}, 067 (2008).

\bibitem{ablPRD2006}
 V.I. Afonso, D. Bazeia, and L. Losano,
     Phys. Lett. \textbf{B 634}, 526 (2006).

\bibitem{bmmPRL2003}
 D. Bazeia, J. Menezes, and R. Menezes,
    Phys. Rev. Lett. \textbf{91,} 241601 (2003).

\bibitem{rs}
 J. Lykken and L. Randall,
       JHEP \textbf{0006}, 014 (2000).

\bibitem{G. Dvali}
G. Dvali and M. Shifman, Phys.Lett. \textbf{B 396}, 64 (1997).

\bibitem{B. Bajc}
B. Bajc and G. Gabadadze, Phys.Lett. \textbf{B 474}, 282 (2000).

\bibitem{S. Randjbar-Daemi}
S. Randjbar-Daemi and M. Shaposhnikov, Phys. Lett. \textbf{B 492},
361 (2000).

\bibitem{S.L. Dubovsky}
S.L. Dubovsky, V.A. Rubakov and P.G. Tinyakov, Phys. Rev.
\textbf{D 62}, 105011 (2000).

\bibitem{A. Melfo}
A. Melfo, N. Pantoja and J.D. Tempo, Phys. Rev. \textbf{D 73},
044033 (2006).

\bibitem{WangMPLA2005}
 Y.Q. Wang, T.Y. Si, Y.X. Liu and Y.S. Duan, Mod. Phys. Lett. \textbf{A 20}, 3045 (2005).

\bibitem{LiuNPB2007}
Y.X. Liu, L. Zhao, X.H. Zhang, Y.S. Duan, Nucl. Phys. \textbf{B
785}, 234 (2007).

\bibitem{LiuVortex}
 Y.X. Liu, Y.Q. Wang and Y.S. Duan, Commun. Theor. Phys. \textbf{48}, 675 (2007).

\bibitem{LiuJHEP2007}
 Y.X. Liu, L. Zhao, Y.S. Duan, JHEP \textbf{0704}, 097 (2007).

\bibitem{DuanGauge}
 Y.S. Duan, Y.X. Liu and Y.Q. Wang, Mod. Phys. Lett. \textbf{A 21}, 2019 (2006).

\bibitem{Mario}
 G. de Pol, H. Singh and M. Tonin, Int. J. Mod. Phys. \textbf{A 15}, 4447 (2000).

\bibitem{JackiwRossiNPB1981}
 R. Jackiw and P. Rossi, Nucl. Phys. \textbf{B 190}, 681 (1981).

\end{thebibliography}
\end{document}